\documentclass[%
 reprint,
superscriptaddress,
showpacs,
 amsmath,amssymb,
 aps,
prc
]{revtex4-1}

\usepackage{graphicx}% Include figure files
\usepackage{dcolumn}% Align table columns on decimal point
\usepackage{bm}% bold math
\usepackage{amsmath}
\usepackage{changes}
\allowdisplaybreaks[1]
\begin{document}

%\preprint{APS/123-QED}

\title{Extended Skyrme interactions for transport model simulations of heavy-ion collisions}% Force line breaks with \\

\author{Rui Wang}
\affiliation{School of Physics and Astronomy and Shanghai Key Laboratory for
Particle Physics and Cosmology, Shanghai Jiao Tong University, Shanghai $200240$, China}
\author{Lie-Wen Chen}
\email[Corresponding author: ]{lwchen@sjtu.edu.cn}
\affiliation{School of Physics and Astronomy and Shanghai Key Laboratory for
Particle Physics and Cosmology, Shanghai Jiao Tong University, Shanghai $200240$, China}
\author{Ying Zhou}
\affiliation{School of Physics and Astronomy and Shanghai Key Laboratory for
Particle Physics and Cosmology, Shanghai Jiao Tong University, Shanghai $200240$, China}

\date{\today}% It is always \today, today,

\begin{abstract}
Based on an extended Skyrme interaction that includes the terms in relative momenta up to sixth order, corresponding to the so-called Skyrme pseudopotential up to next-to-next-to-next-to leading order (N3LO), we derive the expressions of Hamiltonian density and single nucleon potential under general non-equilibrium conditions which can be applied in transport model simulations of heavy-ion collisions induced by neutron-rich nuclei.
While the conventional Skyrme interactions, which include the terms in relative momenta up to second order, predict an incorrect behavior as a function of energy for nucleon optical potential in nuclear matter, the present extended N3LO Skyrme interaction can give a nice description for the empirical nucleon optical potential.
We also construct three interaction sets with different high-density behaviors of the symmetry energy, by fitting both the empirical nucleon optical potential up to energy of $1$ GeV and the empirical properties of isospin asymmetric nuclear matter.
These extended N3LO Skyrme interactions will be useful in transport model simulations of heavy-ion collisions induced by neutron-rich nuclei at intermediate and high energies, and they can also be useful in nuclear structure studies within the mean-field model.
\end{abstract}

%\pacs{21.65.Ef, 21.10.Dr, 26.30.Hj, 21.60.Jz}

\maketitle

%\tableofcontents

\section{Introduction}

One of primary goals of nuclear physics is to study the in-medium nuclear effective interactions.
These studies help us to understand the properties of neutron-rich nuclear matter at
both sub-saturation and supra-saturation densities, which are of fundamental importance in nuclear
physics and astrophysics to investigate the properties of various nuclear
systems or nuclear processes~\cite{LiIJMPE7,Dansc298,Latsc304,BarPR410,StePR411,LatPR442,LiPR464,TraIJMPE21,HorJPG41,LiEPJA50,HebARNPS65,BalPPNP91,OerRMP89,LiNPN27,LiPPNP99}, e.g., heavy-ion collisions
induced by neutron-rich nuclei, the properties of nuclei close to the drip lines,
the nucleosynthesis in different astrophysical sites, the structure of compact stars and the explosion mechanism of supernova.
The equation of state~(EOS) is one of basic properties of nuclear matter, and it
is conventionally defined as the binding energy per nucleon.
While the EOS of symmetric nuclear matter has been relatively well constrained, even up to about five times nuclear saturation density~($\rho_0$) by analyzing the experimental data on giant resonances of finite nuclei~\cite{YouPRL82,LiPRL99,PatPLB726,GarPPNP101} as well as the collective flows and kaon production in heavy-ion collisions~\cite{AicPRL55,Dansc298,FucPPNP56}, the isospin dependent part of the EOS of isospin asymmetric nuclear matter,
which is described by the symmetry energy, is still largely uncertain, especially for its
supra-saturation density behaviors~(see, e.g., Ref.~\cite{LiEPJA50,ChenEPJWC88}).
Determining the density dependence of the symmetry energy provides the main motivation of
many radioactive beam facilities around the world, such as CSR/Lanzhou and BRIF-II/Beijing
in China, SPIRAL2/GANIL in France, FAIR/GSI in Germany, SPES/LNL in Italy, RIBF/RIKEN in Japan,
RAON in Korea, and FRIB/NSCL and T-REX/TAMU in USA.

At sub-saturation densities, information on the EOS of neutron-rich nuclear matter or the symmetry
energy can be obtained from analyzing experimental data either from finite nuclei, e.g., the binding energy,
the charge radius as well as the isovector modes of resonances~\cite{ZZPLB726,BroPRL111,ZZPRC90,ZZPRC92,RocPPNP101},
or from heavy-ion collisions~(HICs) induced by neutron-rich nuclei at intermediate
energies, such as collective flows and particle production~\cite{LiPRL85,TsaPRL92,ChenPRL94,BarPR410,LiPR464,TsaPRL102}.
As for the exploration of the supra-saturation density behaviors of the symmetry energy,
heavy-ion collisions at intermediate to high energies turn out to be a unique tool in terrestrial labs~\cite{LiPRL88,XiaPRL102,FenPLB683,RusPLB697,XiePLB718,CozPRC88,RusPRC94,CozEPJA54}.
Theoretically, microscopic transport models, e.g., the Boltzmann-Uehling-Uhlenbeck~(BUU) equation~\cite{BerPR160}
and the quantum molecular dynamics~(QMD) model~\cite{AicPR202}, provide a powerful tool to extract information
on nuclear matter EOS from analyzing data in heavy-ion collisions~\cite{JXPRC93,ZhaPRC97}.
The microscopic transport models can be also applied to describe some dynamical properties of finite nuclei,
e.g., the giant or pygmy resonances~\cite{YilPRC72,GaiPRC81,UrbPRC85,BarPRC88,BarEPJD68,ZhePRC94,KongPRC95}.
The basic input in one-body transport model (e.g., the BUU equation) is the single nucleon potential~(nuclear mean-field potential) under non-equilibrium conditions since the reaction system in the dynamical process of heavy-ion collisions is generally far from equilibrium.
Due to the exchange term of finite-range nuclear interaction, intrinsic momentum dependence of nuclear interaction, nuclear short-range correlations and other possible contributions, the single nucleon potentials
are generally dependent on nucleon momentum~\cite{BehJPG5,DecPRC21,WirPRC38}, and this is also evident from the observed momentum/energy
dependence of nucleon optical model potential.
In the past few decades, many momentum dependent mean-field potentials have been constructed and developed,
and extensively employed to study both nuclear matter and
heavy-ion collisions~\cite{GalPRC35,PraPRC37,WelPRC38,GrePRC59,DanNPA673,Bom2001,PerPRC65,DasPRC67,ChenPRL94,XuPRC81,JXPRC82,ChenEPJA50,JXPRC91}.

Most of momentum dependent mean-field potentials so far applied in transport model
simulations for heavy-ion collisions are parameterized phenomenologically and they are
difficult to be directly used in mean-field calculations for finite nuclei.
Therefore, it is interesting and constructive to use the same effective interaction to describe the
properties of both finite nuclei and heavy-ion collisions on the same footing.
By doing this, experimental observables from finite nuclei and heavy-ion collisions will
provide crosschecks for the single nucleon potentials and thus obtain more reliable
information on the in-medium nuclear effective interactions and the associated
nuclear matter EOS.

The Skyrme interaction~\cite{SkyPM1,SkyNP9} is perhaps the most popular nuclear effective interaction in nuclear
physics, and it has been used very successfully in describing the ground and lowly excited
state properties of finite nuclei in mean-field calculations~\cite{BenRMP75,StoPPNP58} as well as in the study of heavy-ion collisions
at low energies in time-dependent Hartree-Fock (TDHF) calculations~\cite{MarCPC185,NakRMP88}.
Unfortunately, the conventional Skyrme interactions~\cite{VauPRC5,VauPRC7,ChaNPA627,ChaNPA635} (and their
various variants, see e.g. Ref.~\cite{ChaPRC80,ZZPRC94}), which include only the terms in relative momenta up to second order,
cannot be applied in transport model simulations of heavy-ion collisions at higher
energies~(above about $300$ MeV/nucleon) since the predicted simple parabolic form as a
function of momentum for nucleon mean-field potential fails to reproduce the empirical
results on the nucleon optical potential obtained by Hama \textsl{et al.}~\cite{HamPRC41,CooPRC47}.
This hinders the application of the Skyrme interaction in transport model simulations
of heavy-ion collisions at intermediate and high energies.

In the present work, we demonstrate that the recently developed effective Skyrme pseudopotential~\cite{CarPRC78,RaiPRC83},
which includes additional higher-order derivative terms~(higher-power momentum dependence)
in the conventional Skyrme interactions, may overcome the shortcoming of the conventional
Skyrme interactions and can give a nice description for the empirical nucleon optical
potential up to energy of $1$ GeV. This provides the possibility to study structure properties of finite nuclei and
heavy-ion collisions at incident energy up to about $1$ GeV/nucleon (where the nuclear matter
with about $3\rho_0$ can be formed during the collisions~\cite{LiNPA708})
on the same footing by using the same nuclear effective interaction.
In particular, based on the Skyrme pseudopotential up to next-to-next-to-next-to leading order (N$3$LO) that includes the terms in relative momenta up to sixth order, we derive the Hamiltonian density and isospin- and momentum-dependent single nucleon potential under general non-equilibrium
conditions which can be applied in one-body transport model simulations of heavy-ion collisions
induced by neutron-rich nuclei.
Furthermore, three new parameter sets are obtained by considering the empirical properties
on both EOS of asymmetric nuclear matter and single nucleon potential up to energy of $1$ GeV.

The paper is organized as follows:
In Sec.~\ref{Sec:SkyN3LO}, we introduce Skyrme pseudopotential up to N3LO, and derive
the expressions of the Hamiltonian density and isospin- and momentum-dependent single
nucleon potential under general non-equilibrium conditions.
In Sec.~\ref{Sec:Fit}, we present the experimental data and constraints adopted in our fitting
and give three new interaction parameter sets of the extended Skyrme interactions.
The properties of cold nuclear matter,
including the EOS and single particle behaviors of the newly constructed interactions, are presented in Sec.~\ref{Sec:Result}.
Finally, we summarize our conclusions and make a brief outlook in Sec~\ref{Sec:Summary}.
We present in Appendix~\ref{App:HD} some details of the derivation for
the Hamiltonian density with the extended Skyrme interaction used in the present work.

\section{\label{Sec:SkyN3LO}Theoretical framework}

\subsection{N3LO Skyrme pseudopotential}

Effective interactions with quasi-local operators depending on spatial derivatives
are conventionally called as pseudopotential.
In pervious literatures~\cite{CarPRC78,RaiPRC83}, the Skyrme interaction has been recognized
as the pseudopotential with which a quasi-local nuclear energy density functional~(EDF)
can be derived when averaged within the Hartree-Fock~(HF) approximation, and
a mapping has been established from the N$3$LO local EDF~\cite{CarPRC78} to Skyrme
interaction with additional $4$th and $6$th-order derivative terms.
This is rather important since the quasi-local EDF based on the density-matrix expansion
provides an efficient way to investigate the universal EDF of nuclear system.
On the other hand, the similar pseudopotential containing derivative terms have also been
developed perturbatively up to N$3$LO~($6$th-order derivative terms) based on harmonic-oscillator
effective operators~\cite{HaxPRC77}.
Since the precise structure of nuclear EDF can be derived from low-energy quantum
chromodynamics~(QCD) with chiral perturbation theory~\cite{PugNPA723,KaiNPA724,FinNPA770},
such a mapping from quasi-local EDF to quasi-local effective interaction~(Skyrme interaction)
provides an order-by-order way to examine the validity of each term in the quasi-local
effective interaction.
The generalization of Skyrme interaction with higher-order derivative terms, or usually
called as Skyrme pseudopotential in previous literatures~\cite{CarPRC78,RaiPRC83}, has been
employed to describe EOS of nuclear matter~\cite{DavJPG40,DavJPG41,DavPS90,DavPRC91,DavAA585},
as well as the properties of finite nuclei~\cite{CarCPC181,BecPRC96} ever since it was introduced.

The N3LO Skyrme pseudopotential~\cite{CarPRC78,RaiPRC83} is a generalization of
the standard Skyrme interaction by adding terms that depend on derivative operator~(momentum operator)
up to sixth order, corresponding to the expansion of the momentum space matrix elements of a generic
interaction in powers of the relative momenta up to the sixth order.
In this sense, the standard Skyrme interaction~\cite{VauPRC5,VauPRC7,ChaNPA627,ChaNPA635} is an N1LO Skyrme pseudopotential.
The full Skyrme pseudopotential generally contains spin-independent, spin-orbit and tensor components~(see, e.g., Refs.~\cite{CarPRC78,RaiPRC83,DavJPG41,DavPRC91}).
In the present work we only keep the spin-independent component and ignore the last two components since they do not contribute to spin-averaged quantities, on which we are focusing here.
The corresponding Skyrme interaction used in this work is then written as
\begin{equation}\label{E:VSk}
  v_{Sk}=V^{C}_{\text{N3LO}} + V^{DD}_{\text{N1LO}},
\end{equation}
with the central term
\begin{eqnarray}
&&V^{C}_{\text{N3LO}} =  t_0(1+x_0\hat{P}_{\sigma})\notag\\
&+& t^{[2]}_1(1+x^{[2]}_1\hat{P}_{\sigma})\frac{1}{2}(\hat{\vec{k}}'^2 + \hat{\vec{k}}^2) + t^{[2]}_2(1 + x^{[2]}_2\hat{P}_{\sigma})\hat{\vec{k}}'\cdot\hat{\vec{k}}\notag\\
&+& t^{[4]}_1(1+x^{[4]}_1\hat{P}_{\sigma})[\frac{1}{4}(\hat{\vec{k}}'^2+\hat{\vec{k}}^2)^2+(\hat{\vec{k}}'\cdot\hat{\vec{k}})^2]\notag\\
&+& t^{[4]}_2(1+x^{[4]}_2\hat{P}_{\sigma})(\hat{\vec{k}}'\cdot\hat{\vec{k}})(\hat{\vec{k}}'^2+\hat{\vec{k}}^2)\notag\\
&+& t^{[6]}_1(1 + x^{[6]}_1\hat{P}_{\sigma})(\hat{\vec{k}}'^2 + \hat{\vec{k}}^2)[\frac{1}{2}(\hat{\vec{k}}'^2 + \hat{\vec{k}}^2)^2 + 6(\hat{\vec{k}}'\cdot\hat{\vec{k}})^2]\notag\\
&+& t^{[6]}_2(1+x^{[6]}_2\hat{P}_{\sigma})(\hat{\vec{k}}'\cdot\hat{\vec{k}})[3(\hat{\vec{k}}'^2+\hat{\vec{k}}^2)^2+4(\hat{\vec{k}}'\cdot \hat{\vec{k}})^2],
\label{E:VSkC}
\end{eqnarray}
and the density-dependent term
\begin{equation}\label{E:VSkDD}
  V^{DD}_{\text{N1LO}} = \frac{1}{6}t_3(1+x_3\hat{P}_{\sigma})\rho^{\alpha}(\vec{R}).
\end{equation}
In the above expressions, $\hat{P}_{\sigma}$ represents the
spin exchange operator defined as $\hat{P}_{\sigma} = \frac{1}{2}(1 + \hat{\sigma}_1\hat{\sigma}_2)$,
where $\hat{\sigma}_1$ and $\hat{\sigma}_2$ are Pauli matrices acting on first and second state, respectively;
the $\hat{\vec{k}}'$ and $\hat{\vec{k}}$ are derivative operators acting on left and right, respectively,
and they take the conventional form as $-(\hat{\vec{\nabla}}_1-\hat{\vec{\nabla}}_2)/2i$ and
$(\hat{\vec{\nabla}}_1-\hat{\vec{\nabla}}_2)/2i$; and $\vec{R} = (\vec{r}_1+\vec{r}_2)/2$.
In addition, an overall factor $\hat{\delta}(\vec{r}_1-\vec{r}_2)$ should be understood in
Eqs.~(\ref{E:VSkC}) and (\ref{E:VSkDD}) but omitted here for the sake of clarity.
The density-dependent term $V^{DD}_{\text{N1LO}}$
is taken to be exactly the same as in the standard Skyrme interaction~(see, e.g., Refs.~\cite{ChaNPA627,ChaNPA635}), which is introduced
to phenomenologically mimic the effects of many-body interactions.
The $t^{[n]}_i$, $x^{[n]}_i$ ($n=2, 4, 6$ and $i = 1, 2$), $t_0$, $t_3$, $x_0$, $x_3$ and
$\alpha $ are Skyrme parameters, and the total number of these parameters is $17$
for the present effective interaction.

\subsection{Hamiltonian density and single nucleon potential in one-body transport model}

As mentioned earlier, the single nucleon potential (nuclear mean-field potential) is a
basic input in one-body transport model simulations of heavy-ion collisions.
During heavy-ion collision process, the nucleons are generally in the state far from
equilibrium, and in transport models they are described by the phase space distribution
function (Wigner function) $f_{\tau}(\vec{r},\vec{p})$, with $\tau$ $=$ $1$~[or n] for
neutrons and $-1$~[or p] for protons.
When the collision system approaches to equilibrium, the nucleon distribution function
$f_{\tau}(\vec{r},\vec{p})$ becomes the Fermi-Dirac distribution in momentum space.
For the single nucleon potentials $U_{\tau}(\vec{r},\vec{p})$ used in one-body transport model~\cite{BerPR160}
which is to solve the time evolution of $f_{\tau}(\vec{r},\vec{p})$, we thus need to express
$U_{\tau}(\vec{r},\vec{p})$ in terms of $f_{\tau}(\vec{r},\vec{p})$.
Similarly, the Hamiltonian density ${\cal H}(\vec{r})$ of the collision system can be also
expressed in terms of $f_{\tau}(\vec{r},\vec{p})$, and this is important since we need it
to check the energy conservation of the collisions system during the time evolution in the
transport model simulations.
Moreover, the Hamiltonian density ${\cal H}(\vec{r})$ is also a basic input in one-body
transport model within certain frameworks, e.g., the lattice Hamiltonian
Vlasov method~\cite{LenPRC39}.

In the present work, we derive ${\cal H}(\vec{r})$ and $U_{\tau}(\vec{r},\vec{p})$ in HF
approximation with the extended Skyrme interaction, i.e., Eq.~(\ref{E:VSk}).
The expectation value of the total energy of the collision system can be obtained as
\begin{equation}
\begin{split}
 E &= \sum_i\langle i|\frac{p^2}{2m}|i\rangle + \frac{1}{2}\sum_{i,j}\langle ij|v_{Sk}(1 - \hat{P}_M\hat{P}_{\sigma}\hat{P}_{\tau})|ij\rangle\\
 &\equiv\int {\cal H}(\vec{r}){\rm d}^3r,
 \label{E-DEn}
\end{split}
\end{equation}
where $\hat{P}_M$ and $\hat{P}_{\tau}$ are the Majorana and isospin-exchange operators, respectively.
In the present work we focus on the spin-averaged quantities, and omit all the other irrelevant terms in the Hamiltonian density.
It should be pointed out that for the spin-dependent quantities, time-odd and other spin-dependent terms can be important~\cite{EngNPA249,DobPRC52,UmaPRC73,MarPRC74,JXPLB724,XiaPRC89,XiaPLB759}.
For some specific quantities, for example fusion barrier and cross section, even tensor terms should be taken into consideration~\cite{GuoPLB782}.
The Hamiltonian density ${\cal H}(\vec{r})$ of the collision system with the extended Skyrme interaction
used in the present work can be expressed as~(detailed derivation can be found in Appendix~\ref{App:HD})
\begin{equation}\label{E:H}
\begin{split}
    {\cal H}(\vec{r}) & = {\cal H}^{\rm kin}(\vec{r}) + {\cal H}^{\rm loc}(\vec{r})\\
    & + {\cal H}^{\rm MD}(\vec{r}) + {\cal H}^{\rm grad}(\vec{r}) + {\cal H}^{\rm DD}(\vec{r}),
\end{split}
\end{equation}
where ${\cal H}^{\rm kin}(\vec{r})$, ${\cal H}^{\rm loc}(\vec{r})$, ${\cal H}^{\rm MD}(\vec{r})$,
${\cal H}^{\rm grad}(\vec{r})$ and ${\cal H}^{\rm DD}(\vec{r})$ represent the kinetic, local,
momentum-dependent, gradient, and density-dependent terms, respectively.
The kinetic term
\begin{equation}\label{E:Hkin}
    {\cal H}^{\rm kin}(\vec{r}) = \sum_{\tau = n,p}\int d^3p\frac{p^2}{2m_{\tau}}f_{\tau}(\vec{r},\vec{p}),
\end{equation}
and the local term
\begin{equation}\label{E:Hloc}
    {\cal H}^{\rm loc}(\vec{r}) = \frac{t_0}{4}\bigg[(2 + x_0)\rho^2 - (2x_0 + 1)\sum_{\tau = n,p}\rho_{\tau}^2\bigg],
\end{equation}
are the same as that from the conventional Skyrme interaction(see, e.g., Refs.~\cite{ChaNPA627,ChaNPA635}).
The $\rho_{\tau}(\vec{r})$ in the local term is the nucleon density, which is related to
$f_{\tau}(\vec{r},\vec{p})$ through $\rho_{\tau}(\vec{r})$ $=$ $\int f_{\tau}(\vec{r},\vec{p}){\rm d}^3p$,
and the total nucleon density $\rho(\vec{r}) = \rho_{\rm n}(\vec{r})$ $+$ $\rho_{\rm p}(\vec{r})$.

The momentum-dependent term and gradient term contain the contributions from additional
derivative terms in Eq.~(\ref{E:VSkC}). The momentum-dependent term can be expressed as
\begin{equation}\label{E:HMD}
\begin{split}
    {\cal H}^{\rm MD}(\vec{r}) & = \frac{C^{[2]}}{16\hbar^2}{\cal H}_{\rm}^{md[2]}(\vec{r}) + \frac{D^{[2]}}{16\hbar^2}\sum_{\tau = n,p}{\cal H}_{\tau}^{md[2]}(\vec{r})\\
  & + \frac{C^{[4]}}{32\hbar^4}{\cal H}_{\rm}^{md[4]}(\vec{r}) + \frac{D^{[4]}}{32\hbar^4}\sum_{\tau = n,p}{\cal H}_{\tau}^{md[4]}(\vec{r})\\
  & + \frac{C^{[6]}}{16\hbar^6}{\cal H}_{\rm}^{md[6]}(\vec{r}) + \frac{D^{[6]}}{16\hbar^6}\sum_{\tau = n,p}{\cal H}_{\tau}^{md[4]}(\vec{r}),
\end{split}
\end{equation}
where ${\cal H}^{md[n]}(\vec{r})$ and ${\cal H}_{\tau}^{md[n]}(\vec{r})$ are defined as
\begin{eqnarray}
{\cal H}^{md[n]}(\vec{r}) & = & \int d^3pd^3p'(\vec{p} - \vec{p}')^nf(\vec{r},\vec{p})f(\vec{r},\vec{p}'),\\
{\cal H}_{\tau}^{md[n]}(\vec{r}) & = & \int d^3pd^3p'(\vec{p} - \vec{p}')^nf_{\tau}(\vec{r},\vec{p})f_{\tau}(\vec{r},\vec{p}'),
\end{eqnarray}
with $f(\vec{r},\vec{p})$ $=$ $f_n(\vec{r},\vec{p})$ $+$ $f_p(\vec{r},\vec{p})$.
The gradient term is expressed as
\begin{widetext}
\begin{eqnarray}
{\cal H}^{\rm grad}(\vec{r}) & = & \frac{1}{16}E^{[2]}\Big\{2\rho(\vec{r})\nabla^2\rho(\vec{r}) - 2\big[\nabla\rho(\vec{r})\big]^2\Big\} + \frac{1}{16}F^{[2]}\sum_{\tau = n,p}\Big\{2\rho_{\tau}(\vec{r})\nabla^2\rho_{\tau}(\vec{r}) - 2\big[\nabla\rho_{\tau}(\vec{r})\big]^2\Big\}\notag\\
  & + & \frac{1}{32}E^{[4]}\Big\{2\rho(\vec{r})\nabla^4\rho(\vec{r}) - 8\nabla\rho(\vec{r})\nabla^3\rho(\vec{r}) + 6\big[\nabla^2\rho(\vec{r})\big]^2\Big\}\notag\\
  & + & \frac{1}{32}F^{[4]}\sum_{\tau = n,p}\Big\{2\rho_{\tau}(\vec{r})\nabla^4\rho_{\tau}(\vec{r}) - 8\nabla\rho_{\tau}(\vec{r})\nabla^3\rho_{\tau}(\vec{r}) + 6\big[\nabla^2\rho_{\tau}(\vec{r})\big]^2\Big\}\notag\\
  & + & \frac{1}{16}E^{[6]}\Big\{2\rho(\vec{r})\nabla^6\rho(\vec{r}) - 12\nabla\rho(\vec{r})\nabla^5\rho(\vec{r}) + 30\nabla^2\rho(\vec{r})\nabla^4\rho(\vec{r}) - 20\big[\nabla^3\rho(\vec{r})\big]^2\Big\}\notag\\
  & + & \frac{1}{16}F^{[6]}\sum_{\tau = n,p}\Big\{2\rho_{\tau}(\vec{r})\nabla^6\rho_{\tau}(\vec{r}) - 12\nabla\rho_{\tau}(\vec{r})\nabla^5\rho_{\tau}(\vec{r}) +  + 30\nabla^2\rho_{\tau}(\vec{r})\nabla^4\rho_{\tau}(\vec{r}) - 20\big[\nabla^3\rho_{\tau}(\vec{r})\big]^2\Big\}.\label{E:Hgrad}
\end{eqnarray}
\end{widetext}

In above expressions, for convenience, we have recombined the Skyrme parameters
related to the derivative terms in Eq.~(\ref{E:VSkC}), namely, $t_1^{[n]}$, $t_2^{[n]}$,
$x_1^{[n]}$ and $x_2^{[n]}$, into the parameters $C^{[n]}$, $D^{[n]}$, $E^{[n]}$ and
$F^{[n]}$, i.e.,
\begin{eqnarray}
C^{[n]} & = & t_1^{[n]}(2+x_1^{[n]})+t_2^{[n]}(2+x_2^{[n]}),\label{E:Cn}\\
D^{[n]} & = & -t_1^{[n]}(2x_1^{[n]}+1)+t_2^{[n]}(2x_2^{[n]}+1),\label{E:Dn}\\
E^{[n]} & = & \frac{i^n}{2^n}\big[t_1^{[n]}(2+x_1^{[n]}) - t_2^{[n]}(2+x_2^{[n]})\big],\label{E:En}\\
F^{[n]} & = & -\frac{i^n}{2^n}\big[t_1^{[n]}(2x_1^{[n]}+1) + t_2^{[n]}(2x_2^{[n]}+1)\big]\label{E:Fn}.
\end{eqnarray}
Among these parameters, $C^{[n]}$'s and $D^{[n]}$'s are for momentum-dependent terms
while $E^{[n]}$'s and $F^{[n]}$'s are for gradient terms.
The density-dependent term comes from $V_{\rm N1LO}^{DD}$, i.e., Eq.~(\ref{E:VSkDD}), and can be expressed as
\begin{equation}\label{E:HDD}
\begin{split}
    {\cal H}^{\rm DD}(\vec{r}) = \frac{t_3}{24}\bigg[(2 + x_3)\rho^2 - (2x_3 + 1)\sum_{\tau=n,p}\rho_{\tau}^2\bigg]\rho^{\alpha}.
\end{split}
\end{equation}
If we only keep $2$nd-order terms, Eqs.~(\ref{E:HMD}) and (\ref{E:Hgrad}) are
then reduced to the corresponding momentum-dependent and gradient terms of
the Hamiltonian density from the conventional Skyrme interaction~\cite{ChaNPA627,ChaNPA635}, see, e.g., Ref.~\cite{ZhaPLB732} for the momentum-dependent term.

Within the framework of Landau Fermi liquid theory, the single nucleon energy can be calculated by the variation of ${\cal H}(\vec{r})$ with respect
to $f_{\tau}(\vec{r},\vec{p})$.
Since the Hamiltonian density contains density gradient terms, the single nucleon
potential can then be calculated as~\cite{KolPRC95}
\begin{equation}\label{E:vrtn}
U_{\tau}(\vec{r},\vec{p}) = \frac{\delta{\cal H}^{\rm pot}(\vec{r})}{\delta f_{\tau}(\vec{r},\vec{p})} + \sum_n(-1)^n\nabla^n\frac{\delta{\cal H}^{\rm pot}(\vec{r})}{\delta[\nabla^n\rho_{\tau}(\vec{r})]},
\end{equation}
where ${\cal H}^{\rm pot}(\vec{r})$ $=$ ${\cal H}(\vec{r})$ $-$ ${\cal H}^{\rm kin}(\vec{r})$ is the potential energy density.
Substituting Eq.~(\ref{E:H}) into Eq.~(\ref{E:vrtn}), we then obtain the single
nucleon potential of the extended Skyrme interaction used in the present work, i.e.,
\begin{widetext}
\begin{eqnarray}
U_{\tau}(\vec{r},\vec{p}) & = & \frac{1}{2}t_0\big[(2 + x_0)\rho(\vec{r}) - (2x_0 + 1)\rho_{\tau}(\vec{r})\big] + \frac{\alpha}{24}t_3\bigg[(2 + x_3)\rho(\vec{r})^2 - (2x_3 + 1)\sum _{\tau=n,p}\rho_{\tau}(\vec{r})^2\bigg]\rho(\vec{r})^{\alpha - 1}\notag\\
 & + & \frac{1}{12}t_3\big[(2 + x_3)\rho(\vec{r}) - (2x_3 + 1)\rho_{\tau}(\vec{r})\big]\rho(\vec{r})^{\alpha} + \frac{1}{8\hbar^2}C^{[2]}U^{md[2]}(\vec{r},\vec{p}) + \frac{1}{8\hbar^2}D^{[2]}U_{\tau}^{md[2]}(\vec{r},\vec{p})\notag\\
 & + & \frac{1}{16\hbar^2}C^{[4]}U^{md[4]}(\vec{r},\vec{p}) + \frac{1}{16\hbar^2}D^{[4]}U_{\tau}^{md[4]}(\vec{r},\vec{p}) + \frac{1}{8\hbar^2}C^{[6]}U^{md[6]}(\vec{r},\vec{p}) + \frac{1}{8\hbar^2}D^{[6]}U_{\tau}^{md[6]}(\vec{r},\vec{p})\notag\\
 & + & \frac{1}{8}E^{[2]}\nabla^2\rho(\vec{r}) + \frac{1}{8}F^{[2]}\nabla^2\rho_{\tau}(\vec{r}) + \frac{1}{16}E^{[4]}\nabla^4\rho(\vec{r}) + \frac{1}{16}F^{[4]}\nabla^4\rho_{\tau}(\vec{r}) + \frac{1}{8}E^{[6]}\nabla^6\rho(\vec{r}) + \frac{1}{8}F^{[6]}\nabla^6\rho_{\tau}(\vec{r}),\label{E:U}
\end{eqnarray}
\end{widetext}
where the momentum-dependent terms $U^{md[n]}(\vec{r},\vec{p})$ and
$U_{\tau}^{md[n]}(\vec{r},\vec{p})$ are expressed as
\begin{eqnarray}
U^{md[n]}(\vec{r},\vec{p}) & = & \int d^3p'(\vec{p} - \vec{p}')^nf(\vec{r},\vec{p}'),\\
U_{\tau}^{md[n]}(\vec{r},\vec{p}) & = & \int d^3p'(\vec{p} - \vec{p}')^nf_{\tau}(\vec{r},\vec{p}').
\end{eqnarray}

Based on the above analyses, one can see that the Hamiltonian density
${\cal H}(\vec{r})$ and the single nucleon potential $U_{\tau}(\vec{r},\vec{p})$
are explicitly dependent on $f_{\tau}(\vec{r},\vec{p})$ as well as the local densities
and their derivatives.
It should be mentioned that the momentum dependent part of nuclear mean-field potential could have various origins, e.g., the finite-range exchange term in the nucleon-nucleon interaction, the intrinsic momentum dependence, nucleon short-range correlations.
In Skyrme-Hartree-Fock~(SHF) approach, all these different contributions are combined effectively into the momentum-dependent terms of Eq.~(\ref{E:U}), corresponding to the HF approximation of the derivative terms of the Skyrme interaction.

\subsection{Equation of state of cold nuclear matter}
\label{S-EoS}
For static infinite cold nuclear matter at zero temperature, $f_{\tau}(\vec{r},\vec{p})$
becomes a theta function, i.e., $f_{\tau}(\vec{r},\vec{p}) = \frac{2}{(2\pi\hbar)^3}\theta(p_{\tau}^F - |\vec{p}|)$
with $p_{\tau}^F=\hbar(3\pi^2\rho_\tau)^{1/3}$ being the Fermi
momentum of nucleons with isospin $\tau$ in asymmetric nuclear matter,
and the gradient terms in Eq.~(\ref{E:H}) vanishes.
In this case, the Hamiltonian density can be expressed analytically as a function of
nucleon density $\rho$ and isospin asymmetry $\delta$ $=$ $(\rho_n - \rho_p)/\rho$.
For the extend Skyrme interaction in Eq.~(\ref{E:VSk}), the nuclear matter energy
density is obtained as
\begin{eqnarray}
{\cal E}(\rho,\delta) & = & \frac{3}{5}\frac{\hbar^2a^2}{2m}\rho^{5/3}F_{5/3}\notag\\
& + & \frac{1}{8}t_0^{[0]}\big[2(x_0^{[0]} + 2) - (2x_0^{[0]} + 1)F_2\big]\rho^2\notag\\
& + & \frac{1}{48}t_3^{[0]}\big[2(x_3^{[0]} + 2) - (2x_3^{[0]} + 1)F_2\big]\rho^{(\alpha + 2)}\notag\\
& + & \frac{9a^2}{64}\Big[\frac{8}{15}C^{[2]}F_{5/3} + \frac{4}{15}D^{[2]}F_{8/3}\Big]\rho^{8/3}\notag\\
& + & \frac{9a^4}{128}C^{[4]}\Big(\frac{68}{105}F_{7/3} + \frac{4}{15}\delta G_{7/3} + \frac{4}{15}H_{5/3}\Big)\rho^{10/3}\notag\\
& + & \frac{9a^4}{128}\frac{16}{35}D^{[4]}F_{10/3}\rho^{10/3}\notag\\
& + &\frac{9a^6}{64}C^{[6]}\Big(\frac{148}{135}F_{3} + \frac{4}{5}\delta G_{3} + \frac{4}{5}H_{5/3}F_{2/3}\Big)\rho^4\notag\\
& + &\frac{9a^6}{64}\frac{128}{135}D^{[6]}F_{4}\rho^4,\label{E:EDEoS}
\end{eqnarray}
where $a = (\frac{3\pi^2}{2})^{1/3}$, $m$ is nucleon rest mass in vacuum, and
$F_{x}$, $G_{x}$ and $H_{x}$ are defined as
\begin{align}
F_{x} & = [(1 + \delta)^x + (1 - \delta)^x]/2,\notag\\
G_{x} & = [(1 + \delta)^x - (1 - \delta)^x]/2,\notag\\
H_{x} & = [(1 + \delta)(1 - \delta)]^x.\notag
\end{align}

The EOS of cold nuclear matter can be calculated through dividing its
Hamiltonian density by $\rho$, i.e. $E(\rho,\delta)$ $=$ ${\cal E}(\rho,\delta)/\rho$,
and it can be expanded in the isospin asymmetry $\delta $ as
\begin{equation}
E(\rho,\delta) = E_0(\rho) + E_{\rm{sym}}(\rho)\delta^2 + E_{\rm{sym,4}}(\rho)\delta^4 + {\cal O}(\delta^6),
\label{E-EoSE}
\end{equation}
where $E_0(\rho)$ is the EOS of symmetric nuclear matter,
the symmetry energy $E_{\rm{sym}}(\rho)$ and the fourth-order symmetry energy
$E_{\rm{sym,4}}(\rho)$ are expressed as
\begin{eqnarray}
E_{\rm{sym}}(\rho) & = & \frac{1}{2!}\frac{\partial^2E(\rho,\delta)}{\partial\delta^2}\Big|_{\delta = 0},\\
E_{\rm{sym,4}}(\rho) & = & \frac{1}{4!}\frac{\partial^4E(\rho,\delta)}{\partial\delta^4}\Big|_{\delta = 0}.
\end{eqnarray}

Conventionally, some characteristic parameters defined at saturation density $\rho_0$
are introduced to describe the EOS of asymmetric nuclear matter.
Commonly used are the binding energy per
nucleon in symmetric nuclear matter $E_0(\rho_0)$ and the incompressibility coefficient
$K_0$, the symmetry energy magnitude $E_{\rm{sym}}(\rho_0)$ and its density slope parameter $L$.
In particular, $K_0$ and $L$ are conventionally defined as
\begin{equation}
K_0 = 9\rho_0^2\frac{d^2E_0(\rho)}{d\rho^2}\Big|_{\rho = \rho_0},
L = 3\rho_0\frac{dE_{\rm{sym}}(\rho)}{d\rho}\Big|_{\rho = \rho_0}.
\end{equation}
In order to describe the supra-saturation density behaviors of asymmetric nuclear matter, we usually introduce two higher-order characteristic parameters, i.e., the density skewness coefficient $J_0$ of symmetric nuclear matter and the density curvature parameter $K_{\rm sym}$ of the symmetry energy.
They are defined as~\cite{ChenPRC80}
\begin{equation}
J_0 = 27\rho_0^3\frac{d^3E_0(\rho)}{d\rho^3}\Big|_{\rho = \rho_0},
K_{\rm sym} = 9\rho_0^2\frac{dE^2_{\rm{sym}}(\rho)}{d\rho^2}\Big|_{\rho = \rho_0}.
\end{equation}

\subsection{\label{S-SP}Single nucleon potential in cold nuclear matter}

Similarly as in the case of the Hamiltonian density, for static infinite
cold nuclear matter at zero temperature, the single nucleon potential can be expressed analytically as a function of the magnitude
of nucleon momentum $p$ $=$ $|\vec{p}|$, the nucleon density $\rho$ and isospin
asymmetry $\delta$, i.e.,
\begin{widetext}
\begin{eqnarray}
U_{\tau}(p,\rho,\delta) & = & \frac{1}{4}t_0\big[2(x_0 + 2) - (2x_0 + 1)(1 + \tau\delta)\big]\rho
 +  \frac{1}{24}t_3\big[(\alpha + 2)(x_3 + 2) - (2x_3 + 1)(\frac{\alpha}{2}F_2 + 1 + \tau\delta)\big]\rho^{\alpha + 1}\notag\\
& + & \frac{1}{4}C^{[2]}\Big[\frac{1}{3}\frac{k_F^3}{\pi^2}\Big(\frac{p}{\hbar}\Big)^2 + \frac{1}{5}\frac{k_F^5}{\pi^2}F_{5/3}\Big]
 + \frac{1}{8}D^{[2]}\Big[\frac{1}{3}\frac{k_F^3}{\pi^2}\Big(\frac{p}{\hbar}\Big)^2(1 + \tau\delta) + \frac{1}{5}\frac{k_F^5}{\pi^2}(1 + \tau\delta)^{5/3}\Big]\notag\\
& + & \frac{1}{8}C^{[4]}\Big[\frac{1}{3}\frac{k_F^3}{\pi^2}\Big(\frac{p}{\hbar}\Big)^4 + \frac{2}{3}\frac{k_F^5}{\pi^2}\Big(\frac{p}{\hbar}\Big)^2F_{5/3} + \frac{1}{7}\frac{k_F^7}{\pi^2}F_{7/3}\Big]\notag\\
& + & \frac{1}{16}D^{[4]}\Big[\frac{1}{3}\frac{k_F^3}{\pi^2}\Big(\frac{p}{\hbar}\Big)^4(1 + \tau\delta) + \frac{2}{3}\frac{k_F^5}{\pi^2}\Big(\frac{p}{\hbar}\Big)^2(1 + \tau\delta)^{5/3} + \frac{1}{7}\frac{k_F^7}{\pi^2}(1 + \tau\delta)^{7/3}\Big]\notag\\
& + & \frac{1}{4}C^{[6]}\Big[\frac{1}{3}\frac{k_F^3}{\pi^2}\Big(\frac{p}{\hbar}\Big)^6 + \frac{7}{5}\frac{k_F^5}{\pi^2}\Big(\frac{p}{\hbar}\Big)^4F_{5/3} + \frac{k_F^7}{\pi^2}\Big(\frac{p}{\hbar}\Big)^2F_{7/3} + \frac{1}{9}\frac{k_F^9}{\pi^2}F_3\Big]\notag\\
& + & \frac{1}{8}D^{[6]}\Big[\frac{1}{3}\frac{k_F^3}{\pi^2}\Big(\frac{p}{\hbar}\Big)^6(1 + \tau\delta) + \frac{7}{5}\frac{k_F^5}{\pi^2}\Big(\frac{p}{\hbar}\Big)^4(1 + \tau\delta)^{5/3} + \frac{k_F^7}{\pi^2}\Big(\frac{p}{\hbar}\Big)^2(1 + \tau\delta)^{7/3} + \frac{1}{9}\frac{k_F^9}{\pi^2}(1 + \tau\delta)^{3}\Big],\label{PotSP}
\end{eqnarray}
\end{widetext}
where $k_F = (3\pi^2\rho/2)^{1/3}$ is the Fermi momentum of symmetric nuclear matter,
and $\tau$ equals $1$ for neutrons and $-1$ for protons.

The symmetry potential $U_{\rm{sym,i}}(p,\rho)$ is usually introduced to describe
the isospin dependence of single nucleon potential in asymmetric nuclear matter and
it is defined as
\begin{equation}
\begin{split}
U_{\rm sym,i}(p,\rho) & \equiv \frac{1}{i!}\frac{\partial^iU_n(p,\rho,\delta)}{\partial\delta^i}\big|_{\delta = 0}\\
& = \frac{(-1)^i}{i!}\frac{\partial^iU_p(p,\rho,\delta)}{\partial\delta^i}\big|_{\delta = 0}.
\end{split}
\end{equation}
Then the single nucleon potential in asymmetric nuclear matter can be expressed
as a Taylor expansion with respect to $\delta$, i.e.,
\begin{equation}\label{E:PotTE}
U_{\tau}(p,\rho,\delta) = U_0(p,\rho) + \sum_{i = 1,2,\cdots}\frac{1}{i!}U_{\rm sym,i}(p,\rho)(\tau\delta)^i,
\end{equation}
where $U_0(p,\rho)$ is the single nucleon potential in cold symmetric nuclear matter.
Neglecting higher-order terms in Eq.~(\ref{E:PotTE}), and only keeping $U_0(p,\rho)$
and $U_{\rm sym,1}(p,\rho)$ leads to the well-known Lane potential~\cite{LanNP35},
which has been adopted extensively to approximate the isospin dependent single
particle potential $U_{\tau}(p,\rho,\delta)$.
For the extended Skyrme interaction in Eq.~(\ref{E:VSk}), the (first-order) symmetry
potential can be expressed as
\begin{eqnarray}
U_{\rm sym,1}(p,\rho)
& = & -\frac{1}{4}t_0(2x_0 + 1)\rho - \frac{1}{24}t_3(2x_3 + 1)\rho^{\alpha + 1}\notag\\
& + &\frac{D^{[2]}}{8}\Big[\frac{1}{3}\frac{k_F^3}{\pi^2}\Big(\frac{p}{\hbar}\Big)^2 + \frac{1}{3}\frac{k_F^5}{\pi^2}\Big]\notag\\
& + & \frac{D^{[4]}}{16}\Big[\frac{1}{3}\frac{k_F^3}{\pi^2}\Big(\frac{p}{\hbar}\Big)^4 + \frac{10}{9}\frac{k_F^5}{\pi^2}\Big(\frac{p}{\hbar}\Big)^2 + \frac{1}{3}\frac{k_F^7}{\pi^2}\Big]\notag\\
& + & \frac{D^{[6]}}{8}\Big[\frac{1}{3}\frac{k_F^3}{\pi^2}\Big(\frac{p}{\hbar}\Big)^6 + \frac{7}{3}\frac{k_F^5}{\pi^2}\Big(\frac{p}{\hbar}\Big)^4\notag\\ &+&\frac{7}{3}\frac{k_F^7}{\pi^2}\Big(\frac{p}{\hbar}\Big)^2 + \frac{1}{3}\frac{k_F^9}{\pi^2}\Big].
\label{E-Usym1}
\end{eqnarray}

From the single nucleon potential, one can calculate the
nucleon effective mass in asymmetric nuclear matter, i.e.,
\begin{equation}
m_{\tau}^{\ast} = m\Big[1 + \frac{m}{p}\frac{dU_{\tau}(p,\rho,\delta)}{dp}\Big|_{p=p_{\tau}^F}\Big]^{-1}.
\end{equation}
In addition, the isoscalar effective mass $m_s^{\ast}$ and the isovector
effective mass $m_v^{\ast}$ have been extensively used in nuclear physics.
The $m_s^{\ast}$ is the nucleon effective mass in symmetric nuclear matter,
and $m_v^{\ast}$ can be obtained through~(see, e.g., Ref.~\cite{LiPPNP99} and references therein)
\begin{equation}
    \frac{\hbar^2}{2m_{\tau}^{\ast}} = \frac{2\rho_{\tau}}{\rho}\frac{\hbar^2}{2m_s^{\ast}} + \Big(1-\frac{2\rho_{\tau}}{\rho}\Big)\frac{\hbar^2}{2m_v^{\ast}}.
\end{equation}
The $m_s^{\ast}$ and $m_v^{\ast}$ at saturation density $\rho_0$ are
denoted as $m_{s,0}^{\ast}$ and $m_{v,0}^{\ast}$, respectively.

\section{Fitting strategy and new Skyrme interactions}
\label{Sec:Fit}

In the present work, the parameters of the extended Skyrme interaction in Eq.~(\ref{E:VSk})
are determined by optimizing the weighted sum of squared errors for various
well-known properties of cold nuclear matter.
%\begin{equation}
%  \chi^2 = \sum_{i=1}^{N_{\rm{c}}}(\frac{M_i^{\rm{con}} - M_i^{\rm{th}}}{\sigma_i})^2.
%\end{equation}
%where $N_{\rm{c}}$ is the number of constrains, $M_i^{\rm{con}}$ are central value of constrains and $\sigma_i$ are errors of constrains, $M_i^{\rm{th}}$ are the value calculated by Skyrme effective pseudo-potentials.
Since our main motivation is to develop new extended Skyrme interactions that can be applied in the future into both one-body transport model for heavy-ion collisions (with beam energy up to about $1$ GeV/nucleon) and mean-field calculations for finite nuclei, the extended Skyrme interactions obtained in the present work are required to describe reasonably the well-known empirical momentum dependence of single particle potential $U_0(p,\rho_0)$ in symmetric nuclear matter at saturation density with nucleon momentum up to $1.5$ GeV/c~(approximately corresponding to nucleon kinetic energy of $1$~$\rm{GeV}$).
Thus the real part of nucleon optical potential~(Schr$\rm \ddot{o}$dinger equivalent potential) obtained by Hama~\textsl{et al.}~\cite{HamPRC41,CooPRC47} from dirac phenomenology of the nucleon-nucleus scattering data is used in the optimization.
Furthermore, the new extended Skyrme interactions are also required to give a
prediction on the momentum dependence of the $1$st-order symmetry
potential $U_{\rm{sym,1}}(p,\rho)$ at~[and below] saturation density,
consistent with that from microscopic calculations, for example, the
Brueckner-Hartree-Fock~(BHF) calculation~\cite{DalPRC72} and relativistic
impulse approximation~\cite{ChenPRC72,LiPRC74}~(still for nucleon momentum
up to $1.5$ GeV/c).
Since there are actually no practical errors for the above two properties, we choose $3.16~\rm{MeV}$ and $5~\rm{MeV}$ as the error bars for $U_0(p,\rho_0)$ and $U_{\rm{sym,1}}(p,\rho)$, respectively, in the optimization procedure.

In addition, considering the large uncertainty of the supra-saturation
density behaviors of the symmetry energy, three different extended Skyrme interactions with soft, moderate and hard high density behaviors of the symmetry energy are constructed, by adding constrains $E_{\rm sym}(\rho_{\rm h})$ $=$ $0.0\pm1.0~{\rm MeV}$, $40.0\pm1.0~{\rm MeV}$ and $80.0\pm1.0~{\rm MeV}$, respectively, with $\rho_{\rm h}$ $=$ $0.5~{\rm fm}^{-3}$ in the optimization.

\begin{table}[!htb]
\centering
\caption{Parameters of the three new extended Skyrme interactions, namely,
SP$6$s, SP$6$m and SP$6$h. Here the recombination of Skyrme parameters
defined in Eqs.~(\ref{E:Cn}) and (\ref{E:Dn}) are used.}
%\begin{tabular}{c{3.0cm}c{1.6cm}c{1.6cm}c{1.6cm}}
\begin{tabular}{cccc}
\hline\hline
 ~ & \rm{SP6s} & \rm{SP6m} & \rm{SP6h}\\
\hline
 $t_0$~($\rm{MeV\cdot fm}^{3}$) & -1814.64 & -1956.75 & -1675.52\\
 $x_0$ & 0.5400 & 0.2306 & -0.0902\\
 $t_3$~($\rm{MeV\cdot fm}^{3 + 3\alpha}$) & 10796.2 & 11402.9 & 9873.1\\
 $x_3$ & 0.8257 & 0.1996 & -0.4990\\
 $\alpha$ & 0.2923 & 0.2523 & 0.3168\\
 $C^{[2]}$~($\rm{MeV\cdot fm}^{5}$) & 597.877 & 637.195 & 677.884\\
 $D^{[2]}$~($\rm{MeV\cdot fm}^{5}$) & -446.695 & -524.373 & -601.990\\
 $C^{[4]}$~($\rm{MeV\cdot fm}^{7}$) & -26.2027 & -28.5209 & -31.2026\\
 $D^{[4]}$~($\rm{MeV\cdot fm}^{7}$) & 23.2525 & 27.6873 & 32.4607\\
 $C^{[6]}$~($\rm{MeV\cdot fm}^{9}$) & 0.0903 & 0.1000 & 0.1121\\
 $D^{[6]}$~($\rm{MeV\cdot fm}^{9}$) & -0.0896 & -0.1080 & -0.1292\\
\hline\hline
\end{tabular}
\label{T-SPs}
\end{table}

\begin{figure*}[!htp]
\centering
\includegraphics[width=17cm]{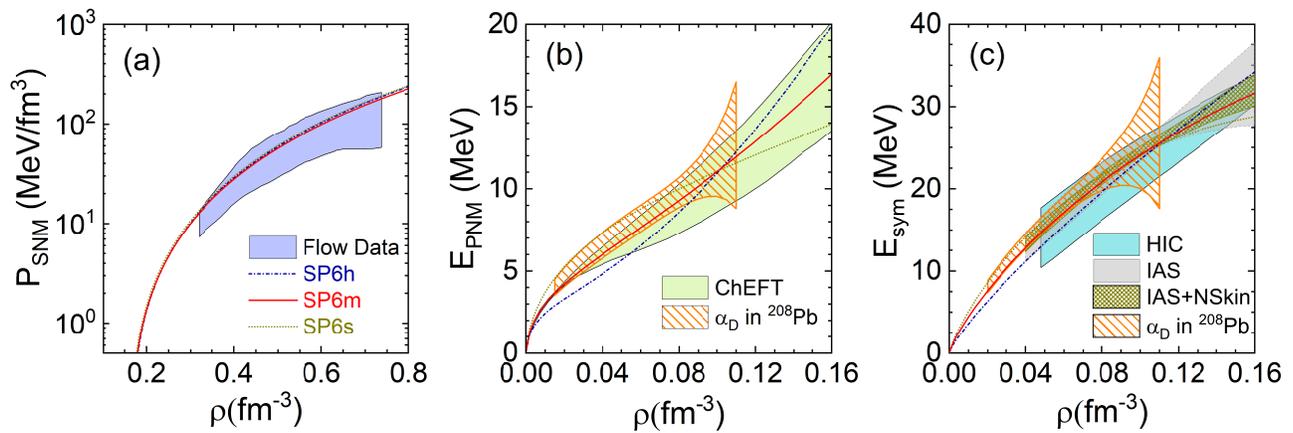}
\caption{Nuclear matter properties with SP6s, SP6m and SP6h, namely,
the density dependence of (a) the pressure of symmetric nuclear matter $P_{\rm{SNM}}(\rho)$,
(b) the EOS of pure neutron matter $E_{\rm{PNM}}(\rho)$ and
(c) symmetry energy $E_{\rm{sym}}(\rho)$.
Also displayed are various constraints on $P_{\rm{SNM}}(\rho)$, $E_{\rm{PNM}}(\rho)$ and
$E_{\rm{sym}}(\rho)$ obtained in previous studies.
See text for details.}
\label{F-NM}
\end{figure*}

Apart from $U_0(p,\rho_0)$, $U_{\rm{sym,1}}(p,\rho)$
and $E_{\rm{sym}}(\rho_{\rm h})$ mentioned above,
the following constrains are also included in the optimization:
(i) the pressure of symmetric nuclear matter in the density region of
$0.3~\rm{fm}^{-3}$ to $0.7~\rm{fm}^{-3}$ should be consistent with
the constraints obtained by analyzing the flow data in HICs~\cite{Dansc298};
(ii) the EOS of pure neutron matter at sub-saturation density
should be consistent with the obtained constrains from electric dipole
polarizability in $^{\rm{208}}\rm{Pb}$~\cite{ZZPRC92};
(iii) the characteristic parameters for symmetric nuclear matter,
$\rho_0$, $E_0$ and $K_0$, are taken to be $0.16\pm0.01~\rm{fm}^{-3}$,
$-16\pm1.0~\rm{MeV}$ and $240\pm30~\rm{MeV}$, respectively;
(iv) we choose an empirical value of $32.5\pm3.2~\rm{MeV}$ as the constrain of $E_{\rm sym}(\rho_0)$ in the optimization.

Since the density gradient terms in the EOS and single nucleon potential
vanish in nuclear matter, the coefficients $E^{[n]}$ and $F^{[n]}$ are
irrelevant in the present fitting with optimization, and the number of the
constrained parameters is thus reduced from $17$ to $11$. To determine the $6$ gradient
coefficients $E^{[n]}$ and $F^{[n]}$ ($n=2,4,6$), one can fit the properties
of finite nuclei, which is beyond the scope of the present work and will be
pursued in future.
In Table~\ref{T-SPs}, we list the values of the $11$ Skyrme parameters for
the obtained three extended Skyrme interaction parameter sets, namely,
SP6s, SP6m, and SP6h, with \emph{s}, \emph{m} and \emph{h}
representing 'soft', 'moderate', and 'hard', respectively, for the
supra-saturation behaviors of the symmetry energy.

\section{\label{Sec:Result}Properties of cold nuclear matter with the new extended Skyrme interactions}

\subsection{Equation of state}

For the three new extended Skyrme parameter sets SP$6$s, SP$6$m
and SP$6$h, the properties of cold nuclear matter can be calculated from
the expressions shown in Section~\ref{Sec:SkyN3LO}, and the results on some
macroscopic characteristic quantities of asymmetric nuclear matter are shown
in Table~\ref{T-CPs}.
It is seen from Table~\ref{T-CPs} that the new extended Skyrme interactions SP$6$s, SP$6$m
and SP$6$h give quite reasonable predictions on the macroscopic characteristic
parameters of asymmetric nuclear matter.
In particular, we include in Table~\ref{T-CPs} the $E_{\rm sym}(\rho_{\rm sc})$ and
$L(\rho_{\rm sc})$ at $\rho_{\rm sc} = 0.11/0.16\rho_0$ which roughly corresponds
to the average density of the heavy nuclei.
The $E_{\rm sym}(\rho_{\rm sc})$ and $L(\rho_{\rm sc})$  are strongly correlated with the isovector properties of finite nuclei and commonly used to characterize the sub-saturation properties of nuclear matter symmetry energy~\cite{ZZPLB726,BroPRL111,ZZPRC90,ZZPRC92}.

\begin{table}[!htb]
\footnotesize
\centering
\caption{Macroscopic characteristic parameters of asymmetric nuclear matter with
SP$6$s, SP$6$m and SP$6$h. Note: $\rho_{\rm sc}$ $=$ $0.11/0.16\rho_0$ and $\rho_{\rm h}$ $=$ $0.5~{\rm fm}^{-3}$.}
%\begin{tabular}{C{2.9cm}C{1.5cm}C{1.5cm}C{1.5cm}}
\begin{tabular}{cccc}
\hline\hline
 ~ & \rm{SP6s} & \rm{SP6m} &\rm{SP6h}\\
 \hline
 $\rho_{0}$~($\rm{fm}^{-3}$) & 0.1614 & 0.1630 & 0.1647\\
 $E_{0}$~($\rm{MeV}$) & -16.04 & -15.94 & -15.61\\
 $K_{0}$~($\rm{MeV}$) & 240.9 & 233.4 & 240.8\\
 $J_0$~($\rm{MeV}$) & $-375.99$ & $-384.16$ & $-358.15$\\
 $E_{\rm{sym}}(\rho_{\rm{sc}})$~($\rm{MeV}$) & 25.43 & 25.83 & 25.98\\
 $L(\rho_{\rm{sc}})$~($\rm{MeV}$) & 32.47 & 46.75 & 62.19\\
 $E_{\rm{sym}}(\rho_0)$~($\rm{MeV}$) & 28.84 & 31.93 & 34.97\\
 $L(\rho_0)$~($\rm{MeV}$) & 18.20 & 49.10 & 82.17\\
  $K_{\rm{sym}}$~($\rm{MeV}$) & $-242.69$ & $-157.98$ & $-70.46$\\
 $E_{\rm{sym}}(\rho_{\rm{h}})$~($\rm{MeV}$) & 0.03 & 41.317 & 79.82\\
 $m_{s,0}^{\ast}/m$ & 0.759 & 0.758 & 0.755\\
 $m_{v,0}^{\ast}/m$ & 0.678 & 0.663 & 0.648\\
\hline\hline
\end{tabular}
\label{T-CPs}
\end{table}

Shown in Fig.~\ref{F-NM} is the density dependence of the pressure of symmetric
nuclear matter $P_{\rm{SNM}}(\rho)$, the EOS of pure neutron matter $E_{\rm{PNM}}(\rho)$,
as well as the symmetry energy $E_{\rm{sym}}(\rho)$ with SP$6$s, SP$6$m
and SP$6$h.
%The corresponding results from three widely used conventional Skyrme interaction sets, namely, SLy$4$~\cite{ChaNPA635}, SKM$^*$~\cite{BarNPA386} and MSL$1$~\cite{ZZPLB726} are also included in Fig.~\ref{F-NM} for comparison.
Also included in Fig.~\ref{F-NM} are the constraints on the $P_{\rm{SNM}}(\rho)$ in
the density region from $2\rho_0$ to $4.6\rho_0$ obtained from analyzing the flow data in
relativistic heavy-ion collisions~\cite{Dansc298},
the $E_{\rm{sym}}(\rho)$ at sub-saturation densities obtained from transport model
analyses of mid-peripheral heavy-ion collisions of Sn isotopes~\cite{TsaPRL102} and
from the SHF analyses of isobaric analog states with or without the combination of
neutron skin data~\cite{DanNPA922},
as well as the constraints on $E_{\rm{PNM}}(\rho)$ and $E_{\rm{sym}}(\rho)$ at
sub-saturation densities recently extracted from the analyses of the electric dipole
polarizability $\alpha_{\rm{D}}$ in $^{208}\rm{Pb}$~\cite{ZZPRC92}.
The result of $E_{\rm{PNM}}(\rho)$ at sub-saturation density calculated in the
framework of chiral effective field theory using N$3$LO potential~\cite{TewPRL110}
is also displayed in Fig.~\ref{F-NM}(b) for comparison.

It should be mentioned that due to the limitation of the density
functional form used in the present work, if one wants to give a correct
single particle behaviors, it is difficult for the parameter set with stiff supra-saturation
symmetry energy, namely, SP$6$h, to give enough soft
sub-saturation symmetry energy, and thus shows small discrepancy compared
to the constrain from electric dipole polarizability $\alpha_{\rm{D}}$
in $^{208}\rm{Pb}$ in Fig.~\ref{F-NM}(b) and Fig.~\ref{F-NM}(c).
However, this small discrepancy will not affect the study of supra-saturation
behavior of the symmetry energy.

\begin{figure}[!htb]
\centering
\includegraphics[width=8.5cm]{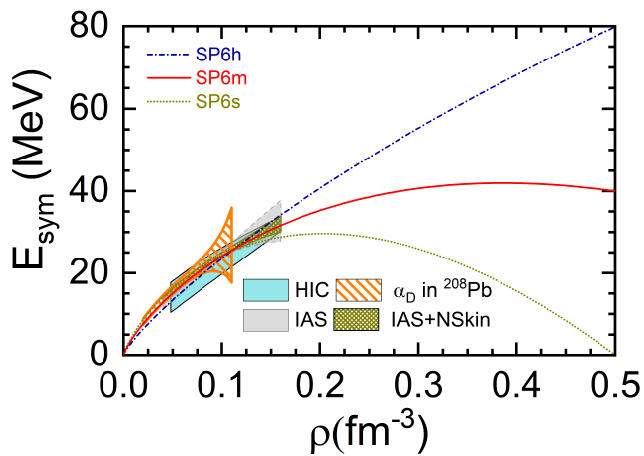}
\caption{Same as Fig.~\ref{F-NM}(c) but extended to supra-saturation density.}
\label{F-Esym}
\end{figure}

To see more clearly the high density behaviors of the symmetry energy from
the three new extended Skyrme interactions, we show in Fig.~\ref{F-Esym} the same
as in Fig.~\ref{F-NM}(c) but extended to the high density region.
One can see from Fig.~\ref{F-Esym} that the symmetry energy from these three new
extended Skyrme interactions exhibit quite different supra-saturation behaviors,
which allows us to extract information on the supra-saturation density behaviors of the symmetry
energy by transport model simulations of heavy-ion collision at intermediate and
high energies, e.g., with beam energy up to about $1$ GeV/nucleon, where the maximum
baryon density of about $3\rho_0$ can be reached~\cite{LiNPA708}.

\subsection{Single nucleon potential}

\begin{figure}[!hbt]
\centering
\includegraphics[width=8.5cm]{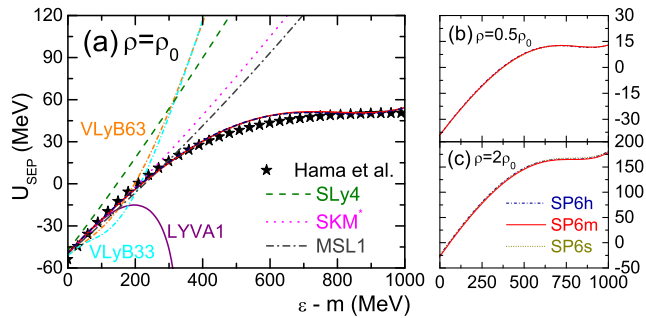}
\caption{The predicted single nucleon potentials in cold symmetric nuclear matter as a
function of nucleon kinetic energy with SP$6$s, SP$6$m and SP$6$h.
The result from three conventional Skyrme interaction, and three N$3$LO Skyrme pseudopotentials obtained in
previous literature, as well as the nucleon optical potential
(Schrodinger equivalent potential) in symmetric nuclear matter at saturation density
$\rho_0$ obtained by Hama \textsl{et al.}~\cite{HamPRC41,CooPRC47} are also included for comparison.
See text for details.}
\label{F-SPP}
\end{figure}

One important improvement of the present extended Skyrme interactions is about the
momentum/energy dependence of the single nucleon potential in nuclear matter.
Shown in Fig.~\ref{F-SPP} is the single nucleon potential $U_0(p,\rho)$ in
cold symmetric nuclear matter at $\rho = \rho_0$, $0.5\rho_0$ and $2\rho_0$,
as a function of nucleon kinetic energy $\epsilon - m = \sqrt{p^2 + m^2} + U_0(p,\rho) - m$
with SP$6$s, SP$6$m and SP$6$h.
Also included in Fig.~\ref{F-SPP} (a) is the real part of nucleon optical potential
(Schrodinger equivalent potential) in symmetric nuclear matter at saturation density
$\rho_0$ obtained by Hama \textsl{et al.}~\cite{HamPRC41,CooPRC47} from Dirac phenomenology of the nucleon-nucleus
scattering data.
In addition, for comparison, we also include in Fig.~\ref{F-SPP} (a)
the corresponding results from three conventional Skyrme interactions SLy$4$~\cite{ChaNPA635},
SKM$^*$~\cite{BarNPA386} and MSL$1$~\cite{ZZPLB726}, as well as the existing N$3$LO Skyrme
pseudopotentials, namely, VLyB$33$ and VLyB$63$~\cite{DavPRC91} and LYVA1~\cite{DavAA585}.

From Fig.~\ref{F-SPP}(a), one can see that at nuclear saturation density,
the new extended Skyrme interactions SP$6$s, SP$6$m and
SP$6$h give a nice description on the empirical nucleon optical potential obtained by
Hama~\textsl{et al.}~\cite{HamPRC41,CooPRC47} for nucleon kinetic energy up to $1$ GeV.
Previous studies on the N3LO Skyrme pseudopotential did not take the single
nucleon potential into consideration, and thus the single particle potential from the existing N$3$LO Skyrme
pseudopotentials VLyB$63$, VLyB$33$~\cite{DavPRC91} and LYVA1~\cite{DavAA585},
show large discrepancy compared to the empirical nucleon optical potential,
especially at nucleon kinetic energies
above about $200$ MeV.
For the three conventional Skyrme interactions SLy$4$~\cite{ChaNPA635},
SKM$^*$~\cite{BarNPA386} and MSL$1$~\cite{ZZPLB726}, due to their simple parabolic
momentum dependence of the single particle potential, they can only give rise
to a reasonable single nucleon potential at lower energies
region (less than about $300$ MeV).
In addition, it is seen that SP$6$s, SP$6$m and SP$6$h predict very
similar single nucleon potential in symmetric nuclear matter at $\rho = 0.5\rho_0$
and $\rho = 2\rho_0$, as shown in Fig.~\ref{F-SPP}(b) and (c), which means
their isospin independent (isoscalar) single particle behaviors are similar
in the relevant density regions interested in the present work.
Such improvement for the extended Skyrme interactions SP$6$s, SP$6$m and SP$6$h obtained in this work allows us to simulate heavy-ion collisions
in one-body transport model for incident energy up to about $1$ GeV/nucleon based
on the Skyrme interaction.
It should be pointed out that the empirical nucleon optical potential obtained by
Hama~\textsl{et al.}~\cite{HamPRC41,CooPRC47} exhibits a clear saturation behavior 
at higher nucleon kinetic energies (i.e., above about $800$ MeV), 
as shown in Fig.~\ref{F-SPP}(a). 
Since the momentum dependence of the extended Skyrme interactions in the present 
work follows polynomial behavior, these interactions may give unsaturated high 
energy mean-field potential and thus are only suitable for describing nuclear 
mean-field potentials in the energy region of less than about $1~\rm{GeV}$.
For higher energy region, 
careful treatments are necessary in dealing with the real part of nucleon optical 
potential, because in the realistic heavy-ion collisions, the nucleon energy could 
be higher than the beam energy due to nucleon-nucleon collisions and Fermi motion.

\begin{figure*}[!hbt]
\centering
\includegraphics[width=16.5cm]{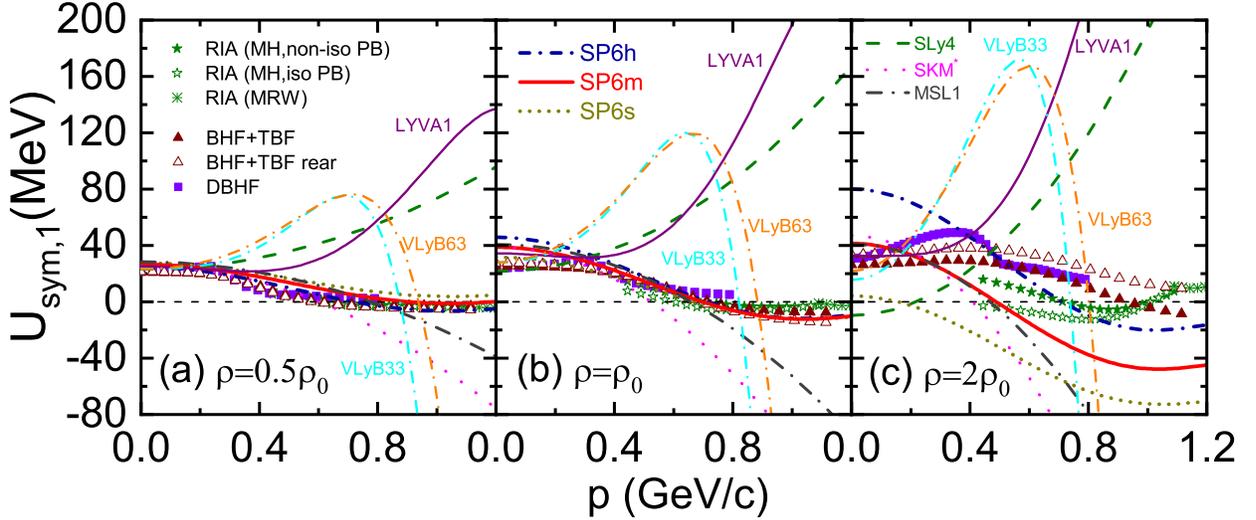}
\caption{The predicted first-order symmetry potential as a function of nucleon momentum
with SP$6$s, SP$6$m and SP$6$h.
The corresponding results from three conventional Skyrme interactions and three existing
N$3$LO Skyrme pseudopotentials, as well as microscopic results from relativistic impulse
approximation and BHF method are also included for comparison.}
\label{F-Usym1}
\end{figure*}

Shown in Fig.~\ref{F-Usym1} is the momentum dependence of the first-order symmetry
potential $U_{\rm sym,1}(p,\rho)$ in cold nuclear matter at $\rho=0.5\rho_0$,
$\rho_0$ and $2\rho_0$, respectively, with SP$6$s, SP$6$m and
SP$6$h.
Also included for comparison are the corresponding results from several microscopic
calculations, namely, the non-relativistic BHF theory with and without rearrangement
contribution from the three-body force~\cite{ZuoPRC74}, the relativistic
Dirac-BHF theory~\cite{DalPRC72}, and the relativistic impulse
approximation~\cite{ChenPRC72,LiPRC74} using the empirical nucleon-nucleon scattering
amplitude determined in Refs.~\cite{MurPRC35,McNPRC27} with
isospin-dependent and isospin-independent Pauli blocking corrections.
The corresponding results from three conventional Skyrme interactions SLy$4$~\cite{ChaNPA635},
SKM$^*$~\cite{BarNPA386} and MSL$1$~\cite{ZZPLB726}, as well as three existing N$3$LO Skyrme pseudopotentials VLyB$33$, VLyB$63$~\cite{DavPRC91} and LYVA1~\cite{DavAA585},  are also included.

It is seen from Fig.~\ref{F-Usym1} that due to the limitation of the parabolic momentum dependence, the results from three conventional Skyrme interactions SLy$4$, SKM$^*$ and MSL$1$ are not satisfactory,
especially at high momenta.
The results with previous N$3$LO Skyrme pseudopotential, VLyB$33$, VLyB$63$ and LYVA$1$ are inconsistent with microscopic calculations as well.
Furthermore, at $\rho=0.5\rho_0$ and $\rho=\rho_0$, the new extended Skyrme interactions SP$6$s, SP$6$m
and SP$6$h predict quite similar $U_{\rm sym,1}(p,\rho)$, and they are
in good agreement with the results of microscopic calculations.
At supra-saturation density of $\rho = 2\rho_0$, the predictions on
$U_{\rm sym,1}(p,\rho)$ from various theoretical approaches display
large discrepancy.

Based on the results shown in Fig.~\ref{F-SPP} and Fig.~\ref{F-Usym1},
we conclude that the new extended Skyrme interactions SP$6$s, SP$6$m and SP$6$h exhibit quite similar
single particle behaviors in asymmetric nuclear matter, except the
isospin dependent (isovector) properties in high density region, where
the microscopic calculations also show large uncertainties, as shown in
Fig. \ref{F-Usym1}(c).
Such differences on the isovector properties of SP$6$s, SP$6$m and SP$6$h at supra-saturation
densities are actually related to their different supra-saturation
behaviors of the symmetry energy depicted in Fig.~\ref{F-Esym}, via the
Hugenholtz-Van Hove theorem~\cite{XuPRC82,RCPRC85,BJCPLB711}.

\section{\label{Sec:Summary}summary and outlook}

Based on an extended Skyrme interaction with extra derivative terms
corresponding to the N$3$LO Skyrme pseudopotential, we have derived
the expressions of Hamiltonian density and single nucleon potentials
for one-body transport models of heavy-ion collisions within the
framework of HF approximation.
We have also obtained
three parameter sets of the extended N3LO Skyrme interaction, by fitting
the empirical properties of cold asymmetric nuclear matter and the single
nucleon optical potential up to energy of $1$ GeV) while
giving soft, moderate and stiff supra-saturation
density behaviors of the symmetry energy, respectively.
Using these extended N3LO Skyrme interactions in one-body transport model to analyze the experimental data in heavy-ion collisions induced by neutron-rich nuclei at intermediate and high energies may help to extract information on the supra-saturation density behavior of nuclear matter EOS, especially the symmetry energy.

The extended N3LO Skyrme interactions can be useful in the study of
both finite nuclei and heavy-ion collisions.
It is interesting to check whether the same Skyrme interaction can
give similar result for nuclear giant resonances within the SHF with random-phase approximation and
the BUU equation, and whether there exist Skyrme interactions
that are able to simultaneously describe the experimental observables
of both finite nuclei and heavy-ion collision.
Such studies are in progress and will be reported elsewhere.

\section*{Acknowledgments}
This work was supported in part by the National Natural Science
Foundation of China under Grant No. 11625521, the Major State Basic Research
Development Program (973 Program) in China under Contract No.
2015CB856904, the Program for Professor of Special Appointment (Eastern
Scholar) at Shanghai Institutions of Higher Learning, Key Laboratory
for Particle Physics, Astrophysics and Cosmology, Ministry of
Education, China, and the Science and Technology Commission of
Shanghai Municipality (11DZ2260700).

\begin{widetext}
\begin{appendix}
\section{\label{App:HD}Hamiltonian density with extended Skyrme interaction in one-body transport model of heavy-ion collisions}

In this appendix, we show some details about the derivation of Hamiltonian density, i.e., Eq.~(\ref{E:H}),
with the extended Skyrme interaction in Eq.~(\ref{E:VSk}).

In quantum mechanics, the Wigner function $f(\vec{r},\vec{p})$ is defined as the
Fourier transform of density matrix and can be thought as the quantum analogy
of classical phase space distribution function in Boltzmann equation.
In coordinate and momentum configuration, $f(\vec{r},\vec{p})$ can be
expressed, respectively, as
\begin{equation}
f(\vec{r},\vec{p}) = \frac{1}{(2\pi\hbar)^3}\int {\rm exp}(-i\frac{\vec{p}}{\hbar}\cdot\vec{s})\rho(\vec{r}+\vec{s}/2,\vec{r}-\vec{s}/2){\rm d}^3s,
\label{E-WT1}
\end{equation}
\begin{equation}
f(\vec{r},\vec{p}) = \frac{1}{(2\pi\hbar)^3}\int {\rm exp}(i\frac{\vec{q}}{\hbar}\cdot\vec{r})g(\vec{p}+\vec{q}/2,\vec{p}-\vec{q}/2){\rm d}^3q,
\label{E-WT2}
\end{equation}
where $\rho(\vec{r}+\vec{s}/2,\vec{r}-\vec{s}/2)$ and $g(\vec{p}+\vec{q}/2,\vec{p}-\vec{q}/2)$
are the density matrix in coordinate and momentum representations, respectively, and
they can be obtained in HF approximation as the matrix elements of one-body density operator
$\hat{\rho}$ $=$ $\sum_i|\phi_{i}\rangle\langle\phi_{i}|$. Specifically, we have
\begin{equation}
  \rho(\vec{r}+\vec{s}/2,\vec{r}-\vec{s}/2) = \langle\vec{r}+\vec{s}/2|\hat{\rho}|\vec{r}-\vec{s}/2\rangle = \sum_i\phi_{i}^{\ast}(\vec{r}+\vec{s}/2)\phi_{i}(\vec{r}-\vec{s}/2),
\end{equation}
in coordinate space and
\begin{equation}
  g(\vec{p}+\vec{q}/2,\vec{p}-\vec{q}/2) = \langle\vec{p}+\vec{q}/2|\hat{\rho}|\vec{p}-\vec{q}/2\rangle = \sum_i\phi_{i}^{\ast}(\vec{p}+\vec{q}/2)\phi_{i}(\vec{p}-\vec{q}/2),
\label{E-DMM} %density matrix in momentum space
\end{equation}
 in momentum space, with $\phi_{i}(\vec{r})$ and $\phi_{i}(\vec{p})$ being the wave functions in
 coordinate and momentum configuration, respectively.
Here for simplicity we ignore the spin index $s$ and isospin index $\tau$.
It can be easily restored by considering the summation of specific state
$|\phi_i\rangle$ with given $s$ and $\tau$.

In the following, we take the $t^{[2]}_1$ term in Eq.~(\ref{E:VSkC})
as a detailed example. The Hamiltonian density can be calculated
through Eq.~(\ref{E-DEn}), and the contribution from the $t^{[2]}_1$
term is
\begin{equation}
 V_1^{[2]} = \frac{1}{2}\sum_{i,j}\langle ij|t^{[2]}_1(1+x^{[2]}_1\hat{P}_{\sigma})\frac{1}{2}(\hat{\vec{k}}'^2\hat{\delta}(\vec{r}_1 - \vec{r}_2)+\hat{\delta}(\vec{r}_1 - \vec{r}_2)\hat{\vec{k}}^2)(1 - \hat{P}_{\sigma}\hat{P}_{\tau}\hat{P}_M)|ij\rangle.
\end{equation}
The parity of this term is positive and thus the Majorana operator $\hat{P}_M$
can be replaced by $\hat{1}$ ($\hat{P}_M$ $=$ $-\hat{1}$ for
negative parity terms).
Assuming that there is no isospin mixing of the HF states, the effect of isospin
exchange operator $\hat{P}_{\tau}$ is just to introduce a Kronecker delta function
$\delta_{\tau_i\tau_j}$ with $\tau_i$ represents the isospin of the $i$-th state.
We assume the collision system is spin saturate since we only focus on the spin-averaged quantities.
For a spin saturation system, the spin exchange operator $\hat{P}_{\sigma}$ is simply replaced by a factor of $\frac{1}{2}$.
Therefore, $V_1^{[2]}$ can be reduced to
\begin{equation}
\begin{split}
 V_1^{[2]} & = \frac{1}{2}t^{[2]}_1(1+\frac{1}{2}x^{[2]}_1)\sum_{i,j}\langle ij|\frac{1}{2}(\hat{\vec{k}}'^2\hat{\delta}(\vec{r}_1 - \vec{r}_2) + \hat{\delta}(\vec{r}_1 - \vec{r}_2)\hat{\vec{k}}^2)|ij\rangle\\
 & - \frac{1}{2}t^{[2]}_1(x^{[2]}_1 + \frac{1}{2})\delta_{\tau_i\tau_j}\sum_{i,j}\langle ij|\frac{1}{2}(\hat{\vec{k}}'^2\hat{\delta}(\vec{r}_1 - \vec{r}_2) + \hat{\delta}(\vec{r}_1 - \vec{r}_2)\hat{\vec{k}}^2)|ij\rangle.
\end{split}
\end{equation}
The calculation of the sum of the bracket is quite straightforward.
It is very convenient if we treat the derivative operator $\nabla$ in $\hat{\vec{k}}'$
and $\hat{\vec{k}}$ as the momentum operator acting on the momentum eigenstate,
and the isospin symmetric part~(similar for the asymmetric part)
can be calculated as
\begin{equation}
\begin{split}
v_1^{[2]} \equiv &~\sum_{i,j}\langle ij|\frac{1}{2}(\hat{\vec{k}}'^2\hat{\delta}(\vec{r}_1 - \vec{r}_2) + \hat{\delta}(\vec{r}_1 - \vec{r}_2)\hat{\vec{k}}^2)|ij\rangle\\
 = &~\frac{1}{4\hbar^2}\sum_{i,j}\int d^3r\int d^3p_1d^3p_2d^3p_3d^3p_4\frac{1}{2}\big[(\vec{p}_1 - \vec{p}_2)^2 + (\vec{p}_3 - \vec{p}_4)^2\big]{\rm exp}\Big[-\frac{i(\vec{p}_1 + \vec{p}_2 - \vec{p}_3 - \vec{p}_4)\cdot\vec{r}}{\hbar}\Big]\\
 &~\times\frac{1}{(2\pi\hbar)^6}\phi_i^{\ast}(\vec{p}_1)\phi_j^{\ast}(\vec{p}_2)\phi_i(\vec{p}_3)\phi_j(\vec{p}_4)\\
 = &~\frac{1}{4\hbar^2}\int d^3r\int d^3pd^3p'd^3qd^3q'\Big[(\vec{p} - \vec{p}')^2 + \Big(\frac{\vec{q} - \vec{q}'}{2}\Big)^2\Big]{\rm exp}\Big(\frac{i\vec{q}\cdot\vec{r}}{\hbar}\Big){\rm exp}\Big(\frac{i\vec{q}'\cdot\vec{r}}{\hbar}\Big)\\
 &~\times\frac{1}{(2\pi\hbar)^6}g_{\vec{p}+\vec{q}/2,\vec{p}-\vec{q}/2}\cdot g_{\vec{p}'+\vec{q}'/2,\vec{p}'-\vec{q}'/2}.
\end{split}
\label{v12}
\end{equation}
To obtain the second line in the above equation we have inserted several unit
operators~(i.e., 4 in coordinate space $\int dr|r\rangle\langle r|$ and 6 in momentum
space $\int dp|p\rangle\langle p|$), and in the last line we have changed the integral
variables to $\vec{p}$ $=$ $\frac{\vec{p}_1 + \vec{p}_3}{2}$,
$\vec{p}'$ $=$ $\frac{\vec{p}_2 + \vec{p}_4}{2}$, $\vec{q}$ $=$ $\vec{p}_3 - \vec{p}_1$
and $\vec{q}'$ $=$ $\vec{p}_4 - \vec{p}_2$.
We have also replaced, as defined in Eq.~(\ref{E-DMM}),
$\sum_i\phi_i^{\ast}(\vec{p}_1)\phi_i(\vec{p}_3)$ and
$\sum_i\phi_i^{\ast}(\vec{p}_2)\phi_i(\vec{p}_4)$ by
$g_{\vec{p}+\vec{q}/2,\vec{p}-\vec{q}/2}$ and
$g_{\vec{p}'+\vec{q}'/2,\vec{p}'-\vec{q}'/2}$, respectively.

The integration of the $(\vec{p} - \vec{p}')^2$ term in Eq.~(\ref{v12}) can be reduced,
through the definition of Wigner function in Eq.~(\ref{E-WT2}), to
\begin{equation}
\frac{1}{4\hbar^2}\int d^3r\int d^3pd^3p'(\vec{p} - \vec{p}')^2f(\vec{r},\vec{p})f(\vec{r},\vec{p}'),
\end{equation}
while the integration of the $(\frac{\vec{q} - \vec{q}'}{2})^2$ term  in Eq.~(\ref{v12})
is recognized as the derivatives of density $\rho(\vec{r})$ through the relation
\begin{equation}
\begin{split}
  \nabla^n\rho(\vec{r}) = & \nabla^n\int f(\vec{r},\vec{p})d^3p = \nabla^n\int d^3p\frac{1}{(2\pi\hbar)^3}\int {\rm exp}(i\frac{\vec{q}}{\hbar}\cdot\vec{r})g_{\vec{p}+\vec{q}/2,\vec{p}-\vec{q}/2}d^3q\\
  = & \int d^3p d^3q\frac{1}{(2\pi\hbar)^3}\Big(i\frac{\vec{q}}{\hbar}\Big)^n{\rm exp}(i\frac{\vec{q}}{\hbar}\cdot\vec{r})g_{\vec{p}+\vec{q}/2,\vec{p}-\vec{q}/2}.
\end{split}
\label{E-DTs}
\end{equation}
In general, derivative terms $\nabla^n\rho(\vec{r})$ will contribute to the
Hamiltonian density and the single nucleon potential and affect the motion
of the nucleons underneath the mean-field potentials, and thus they cannot
be omitted in one-body transport models.

Finally, for $t^{[2]}_1$ term in Eq.~(\ref{E:VSkC}), one can obtain its
contribution to Hamiltonian density as
\begin{equation}
\begin{split}
    V_1^{[2]} & = \frac{1}{16\hbar^2}t_1^{[2]}(2 + x_1^{[2]})\int{\rm d}^3p{\rm d}^3p'(\vec{p} - \vec{p}')^2f(\vec{r},\vec{p})f(\vec{r},\vec{p}')\\
    & - \frac{1}{16\hbar^2}t_1^{[2]}(2x_1^{[2]} + 1)\sum_{\tau = n,p}\int{\rm d}^3p{\rm d}^3p'(\vec{p} - \vec{p}')^2f_{\tau}(\vec{r},\vec{p})f_{\tau}(\vec{r},\vec{p}')\\
    & - \frac{1}{64}t_1^{[2]}(2 + x_1^{[2]})\Big\{2\rho(\vec{r})\nabla^2\rho(\vec{r}) - 2\big[\nabla\rho(\vec{r})\big]^2\Big\} + \frac{1}{64}t_1^{[2]}(2x_1^{[2]} + 1)\sum_{\tau = n,p}\Big\{2\rho_{\tau}(\vec{r})\nabla^2\rho_{\tau}(\vec{r}) - 2\big[\nabla\rho_{\tau}(\vec{r})\big]^2\Big\}.
\end{split}
\end{equation}
The contributions to the Hamiltonian density from other terms
in Eqs.~(\ref{E:VSkC}) and (\ref{E:VSkDD}) can be derived similarly.
After changing the Skyrme parameters $t_1^{[n]}$, $t_2^{[n]}$, $x_1^{[n]}$
and $x_2^{[n]}$ to the parameters in Eqs.~(\ref{E:Cn}) - (\ref{E:Fn}),
one can then obtain the Hamiltonian density Eq.~(\ref{E:H}) of the system
with the extended Skyrme interaction as well as the relevant expressions
in Eqs.~(\ref{E:Hkin}) - (\ref{E:Hgrad}) and Eq.~(\ref{E:HDD}).

\end{appendix}

\end{widetext}
\bibliography{SP-HIC}

\end{document}